\documentclass{emulateapj}
\usepackage{epstopdf}

\newcommand{\goes}{{\it GOES} }
\newcommand{\goess}{{\it GOES}}
\newcommand{\rhessi}{{\it RHESSI} }
\newcommand{\rhessis}{{\it RHESSI}}

\begin{document}

\title{THE THERMAL PROPERTIES OF SOLAR FLARES OVER THREE SOLAR CYCLES USING \goes X-RAY OBSERVATIONS}

\author{Daniel F. Ryan\altaffilmark{1}, Ryan O. Milligan\altaffilmark{2,3,6}, Peter T. Gallagher\altaffilmark{1}, Brian R. Dennis\altaffilmark{2}, A. Kim Tolbert\altaffilmark{2,4}, Richard A. Schwartz\altaffilmark{2,3}, \& C. Alex Young\altaffilmark{2,5}}

\altaffiltext{1}{School of Physics, Trinity College Dublin, Dublin 2, Ireland.}
\altaffiltext{2}{Solar Physics Laboratory (Code 671), Heliophysics Science Division, NASA Goddard Space Flight Center, Greenbelt, MD 20771, U.S.A.}
\altaffiltext{3}{Catholic University of America, Washington, DC 20064, U.S.A.}
\altaffiltext{4}{Wyle Information Systems Inc.}
\altaffiltext{5}{ADNET Systems Inc.}
\altaffiltext{6}{Current address: Queen's University Belfast, Belfast BT7 1NN, Northern Ireland.}

\begin{abstract}
Solar flare X-ray emission results from rapidly increasing temperatures and emission measures in flaring active region loops. To date, observations from the X-Ray Sensor (XRS) onboard the {\it Geostationary Operational Environmental Satellite (GOES)} have been used to derive these properties, but have been limited by a number of factors, including the lack of a consistent background subtraction method capable of being automatically applied to large numbers of flares. In this paper, we describe an automated temperature and emission measure-based background subtraction method (TEBBS), which builds on the methods of \cite{born90}. Our algorithm ensures that the derived temperature is always greater than the instrumental limit and the pre-flare background temperature, and that the temperature and emission measure are increasing during the flare rise phase. Additionally, TEBBS utilizes the improved estimates of \goes temperatures and emission measures from \cite{whit05}. TEBBS was successfully applied to over 50,000 solar flares occurring over nearly three solar cycles (1980-2007), and used to create an extensive catalog of the solar flare thermal properties.  We confirm that the peak emission measure and total radiative losses scale with background subtracted \goes X-ray flux as power-laws, while the peak temperature scales logarithmically. As expected, the peak emission measure shows an increasing trend with peak temperature, although the total radiative losses do not. While these results are comparable to previous studies, we find that flares of a given \goes class have lower peak temperatures and higher peak emission measures than previously reported.  The resulting TEBBS database of thermal flare plasma properties is publicly available on Solar Monitor (www.solarmonitor.org/TEBBS/) and will be available on Heliophysics Integrated Observatory (www.helio-vo.eu).

\end{abstract}

\section{Introduction}
\label{sec:intro}
Solar flares are among the most powerful events in the solar system, releasing up to $10^{33}$ ergs in a few hours or even minutes. They are believed to be powered by magnetic reconnection, a process whereby energy stored in coronal magnetic fields is suddenly released. According to the CSHKP flare model \citep{carm64,stur66,hira74,kopp76}, electrons accelerated by magnetic reconnection spiral down the magnetic loops and strike the chromosphere causing the emission of hard X-rays (HXR). As a consequence, the chromospheric material is also heated and expands back up into the loops which causes the observed increase in temperature and emission measure \citep[e.g.,][]{flet11}.

To date, the study of solar flares has been predominantly focused on single events or small samples of events. While such studies have furthered our understanding of the physics of these particular flares, they are fundamentally limited since they cannot, with any certainty, explain the global behavior of solar flares. In contrast, only the study of large-scale samples can give an insight as to whether findings of given studies are particular to individual events or characteristic of many. This can allow constraints to be placed on global flare properties and give a greater understanding to the fundamental processes which drive these explosive phenomena.

That said, large-scale studies of solar flare properties have been few in number over the past decades. Such a study was performed by \cite{garc92} who used the Geostationary Operational Environmental Satellite (\goess) to examine 710 M- and X-class flares.  They noted a sharp linear lower bound in the relationship between emission measure and \goes class.  However, this paper is mainly focused on categorizing types of very high temperature flares and examined whether these flares approached or exceeded this emission measure lower bound.

A definitive example of a large-scale study of the thermal properties of solar flares was conducted by \cite{feld96b}, who combined results from three previous studies \citep{phil95,feld95,feld96a} to investigate how temperature and emission measure vary with respect to \goes class for 868 flares, from A2 to X2. Their work used temperatures derived using the Bragg Crystal Spectrometer (BCS) onboard \emph{Yohkoh}. These temperature values were convolved with the corresponding \goes data to derive values of emission measure. They found a logarithmic relationship between \goes class and temperature, and a power-law relationship between \goes class and emission measure, with larger flares exhibiting higher temperatures and emission measures. However, temperature and emission measure were derived at the time of the peak 1--8~\AA~flux and so are likely to be less than their true maxima. Furthermore, BCS temperatures have been found to be higher than those measured by \goes \citep{feld96b}, and using these values to calculate \goes emission measure will give lower values than if \goes was used consistently.

More recently, \cite{batt05} studied the correlation between temperature and \goes class for a sample of 85 flares, ranging from B1 to M6 class.  Although the values reported gave a flatter dependence than \cite{feld96b}, the large scatter in the data led to a very large uncertainty making the two relations comparable.  In contrast to \citet{feld96b}, \citet{batt05} accounted for solar background and extracted the flare temperature at the time of the HXR burst as measured by the {\it Ramaty High-Energy Solar Spectroscopic Imager} (\rhessis; \citealt{lin02}) rather than at the time of the soft X-ray (SXR) peak. However, any discrepancies expected to be caused by these differences were not discernible in view of the large uncertainties.  In addition, \citet{casp10} used \rhessi to examine the temperature of 37 high temperature flares with \goes class.  The relationship was qualitatively similar to those of \citet{feld96b} and \citet{batt05} however as the relationship was not fit no quantitative comparison can be made.

Larger statistical samples were studied by \cite{chri08} and \cite{hann08}, who investigated the frequency distributions and energetics of 25,705 microflares (\goes class A--C) observed by \rhessi from 2002 to 2007. From those events for which an adequate background subtraction could be performed (6,740) a median temperature of $\sim$13~MK and emission measure of 3$\times$10$^{46}$~cm$^{-3}$ were found. \cite{hann08}, in particular, looked at the temperature derived from \rhessi observations as a function of (background subtracted) \goes class, and found similar trends to the works of \cite{feld96b} and \cite{batt05}. However, their analysis only included events of low C-class and below.

While these studies have provided some insight into the global properties of solar flares, they each have their limitations. In particular they lack a commonly used method of isolating the flare signal from the solar and instrumental background contributions.  Previous background subtraction methods have often been performed manually. Others, such as setting the background to the flare's initial flux values, or fitting polynomials between the flux values at the start and end of the flare, often exaggerate noise and do not preserve characteristic temperature and emission measure evolution. Therefore, the accurate separation of flare signal and background limits the number of events that can be analyzed. For example, \cite{batt05}, in accounting for solar background, were only able to compile a sample of 85 events.  Although a larger dataset would not have reduced the range of scatter, it would have better revealed the variations in the density of points within the distribution.  This would have allowed a fit to be more tightly constrained and thereby reduced the uncertainties. Conversely, \cite{feld96b}, with a sample of hundreds of flares, did not attempt to account for the solar background at all, which can bias smaller events as the background makes up a greater contribution to the overall flux.

Few attempts have been made to develop automated background subtraction techniques for \goes observations which can be applied to large numbers of flares. \cite{born90} developed a method to determine whether a given background subtraction preserves characteristic temperature and emission measure evolution without checking manually, i.e., that temperature and emission measure both increase during the rise phase of flares.  This behaviour has been seen in numerous observations (e.g., \citealt{flud95,batt09}) and numerical models (e.g., \citealt{fisc85,asch09}). This method used the polynomials of \citet{thom85} which relate temperature and emission measure to the ratio of the short and long \goes channels, $R=F_S/F_L$.  However, \cite{whit05} have since improved on this by assuming more modern spectral models \citep[CHIANTI 4.2][]{land99,land02} and taking into account the differences between coronal and photospheric abundances, requiring the tests of \cite{born90} to be updated.

In this paper, we study the thermal properties of solar flares using \goes observations over nearly three solar cycles.  The flare signal within these observations has been isolated from the various solar, non-solar, and instrumental background contributions using a modified background subtraction method.  This in turn has allowed more accurate automatic calculation of flare properties.  In Section~\ref{sec:goes} of this paper, we discuss the \goes XRS, the \goes event list and how to derive plasma properties from \goes observations.  In Section~\ref{sec:method} we describe previous background subtraction methods for \goes observations and outline how we have improved upon the work of \cite{born90}.  In Section~\ref{sec:results} we use this method to improve upon previous statistical studies by deriving flare properties such as peak temperature and emission measure for flares in the \goes event list and examining the relationships between them.  In Sections~\ref{sec:discussion} and \ref{sec:conclusions} we discuss the results and provide some conclusions.

\section{Observations \& Data Analysis}
\label{sec:goes}
\subsection{Geostationary Operational Environmental Satellite/X-Ray Sensor}
The observations used in this study have been made by satellites of the \goes (Geostationary Operational Environmental Satellite) series, which has been operated by the National Oceanographic and Atmospheric Administration (NOAA) since 1976.  Each spacecraft carries an X-Ray Sensor (XRS) onboard which measures the spatially integrated solar X-ray flux in two wavelength bands (long; 1--8~\AA, and short; 0.5--4~\AA) every three seconds. The sensitivities of the various XRS instruments have remained comparable over the years, although the design of the \goess-8 XRS and subsequent detectors was altered due to the change from a spin-stabilized to 3-axis stabilized platform. For an in-depth discussion of the \goess-8 XRS see \cite{hans96}. The \goes series has provided a near-uninterrupted catalog of solar activity for over three complete solar cycles, and the \goes flare classification scheme is now universally accepted.

\subsection{The \goes Event List}
\label{sec:gev}
The sample of flares used in this study has been extracted from the \goes event list; a list of solar X-ray events which has been compiled by NOAA throughout the lifetime of the \goes series. In order for a solar flare to be included in the \goes event list, it must satisfy two criteria\footnote{\url{http://www.swpc.noaa.gov/ftpdir/indices/events/README}}: firstly, there must be a continuous increase in the one-minute averaged X-ray flux in the long channel for the first four minutes of the event; secondly, the flux in the fourth minute must be at least 1.4 times the initial flux.  The start time of the event is defined as the first of these four minutes. The peak time is when the long channel flux reaches a maximum and the end of an event is defined as the time when the long channel flux reaches a level halfway between the peak value and that at the start of the flare.

\begin{figure}
\begin{center}
\includegraphics[height=9cm,angle=90]{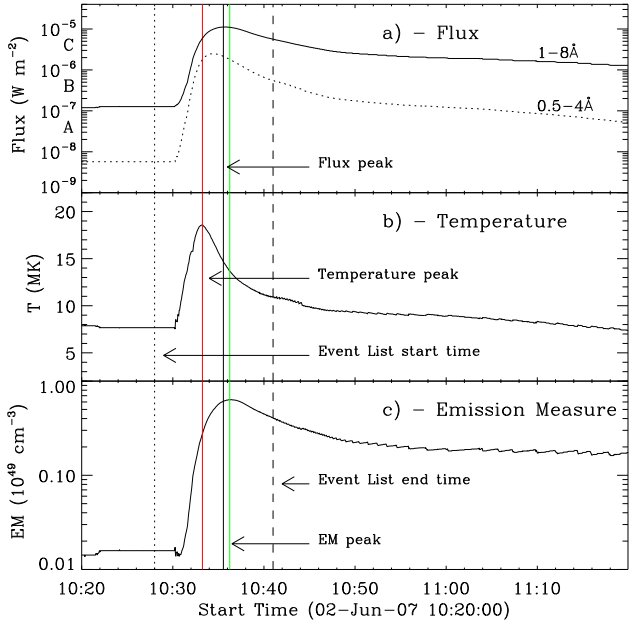}
\caption{X-ray lightcurves of an M1.0 solar flare observed by \goess. {\em a)} X-ray flux in each of the two \goes channels (0.5--4~\AA; dotted curve and 1--8~\AA; solid curve).  {\em b)} The derived temperature curve. {\em c)} The derived emission measure curve. The vertical dotted and dashed lines denote the defined start and end times of the event, respectively. The vertical red, black and green lines mark the times of the peak temperature, peak 1--8~\AA~flux, and peak emission measure, respectively.}
\label{fig:goodprofiles}
\end{center}
\end{figure}

The flare start and end times determined by these definitions do not always agree with those identified manually.    An example of this can be seen in Figure~\ref{fig:goodprofiles}{\it a} which shows the X-ray fluxes in the two \goes channels for an M1.0 solar flare that occurred on 2007 June 2.  The event list start and end times are marked by the vertical dotted and dashed lines, respectively.  The start time of the \goes event is a couple of minutes before the onset of the flare.  Nonetheless, this start time satisfies the event list criteria and highlights a drawback in the event list definitions.  Another drawback is associated with the event list end time.  It can clearly be seen that the decay of the flare in Figure~\ref{fig:goodprofiles}{\it a} continues for over half an hour after the event list end time.  This means that properties depending on the decay time or duration of the flare, such as total radiative losses, will be systematically underestimated.

However, these definitions also help reduce the number of `double-flares' in the event list, i.e., two flares being incorrectly labeled as one.  This can happen when one flare occurs on the decay of another thereby preventing the full-disk integrated flux reaching half the peak value of the first flare.  Having searched the event list between 1991 and 2007 we found 1,865 out of 34,361 events (5.4\%) contained points between their peak and end times which satisfied the event list start criteria.  Of these, the second flare was recorded in the event list in 236 cases.  It should also be stated that the event list start criteria do not locate small events (e.g., B-class) at times of high background flux or during large flares (e.g., M-class).  This is because a small flare will not cause the full-disk integrated X-ray flux to increase to 1.4 times the initial value when that initial value is more than an order of magnitude greater than the flare itself.  Therefore, although one would expect to always find more small events, the event list actually contains fewer around solar maximum when large events are more frequent and the background is often at the C1 level or higher.

The \goes event list for the period 1980 to 2007 was used in this study.  Data from the 1970s were not included due to their poor quality and because many \goes events from this period were erroneously tagged. This meant that a total of 60,424 events, from B-class to X-class, were considered. Events for which data were unavailable, erroneously included (i.e., did not satisfy the event list definitions), or displayed data drop-outs were then removed.  It was found that the size distribution of the discarded events was very similar to that of the entire data set. This implies that the remaining dataset was not biased by the exclusion of these events. After these events were removed, 52,573 remained.

\subsection{Deriving Flare Plasma Parameters}
\label{sec:deriv}
Although \goes only measures X-ray flux in two passbands, techniques have been developed to derive flare plasma properties from the ratio of the short and long channels \citep[e.g.][]{thom85,garc94}. These properties include temperature, $T$, emission measure, $EM$, and the total radiative loss rate from the X-ray emitting plasma, $dL_{rad}/dt$.  In this study, temperature and emission measure were computed from the calculations of \cite{whit05}, who used updated detector responses and plasma source functions to create tables of the dependence of temperature and emission measure on the fluxes in \goes channels.  This method is an updated version of that of \citet{thom85} who derived temperature and emission measure relations from polynomial fits to the \goess-1 XRS response function.  The tabulated values of temperature and emission measure given in \cite{whit05} can be approximated using:
\begin{equation}
T = A_0 + A_1R + A_2R^2 + A_3R^3 \hspace{2.5mm} \mbox{MK}
\label{eqn:temp}
\end{equation}
and
\begin{equation}
EM = F_L \times \frac{1}{B_0 + B_1T + B_2T^2 + B_3T^3} {\rm 10^{49} \mbox{  cm$^{-3}$}}
\label{eqn:em}
\end{equation}
where $R (= F_S/F_L)$ is the ratio of the short to long channel.  For values of the coefficients $A_n$ and $B_n$ for each \goes satellite, see Table 2 of \cite{whit05}.  These tables cover the range from 1--100~MK.  However, due to instrumental sensitivities of the XRS instruments used in this study they are only valid above 4~MK.

Figures~\ref{fig:goodprofiles}{\it b} and \ref{fig:goodprofiles}{\it c} show the temperature and emission measure evolution of the 2007 June 2 flare. Their behavior is {\it characteristic} of the evolution of a typical flare. According to the standard flare model  \citep[][CSHKP]{carm64,stur66,hira74,kopp76}, nonthermal electrons rapidly heat the chromosphere to high temperatures (Figure~\ref{fig:goodprofiles}{\it b}), causing the plasma to expand into overlying flare loops in a process called chromospheric evaporation. These high loop densities result in an increased emission measure following the temperature peak (Figure~\ref{fig:goodprofiles}{\it c}). Once heating has ceased, the plasma cools by thermal conduction and then via radiative processes. This is accompanied by a progressive decrease in both flare temperature and emission measure. 

Theoretically, the radiative loss rate, $dL_{rad}/dt$, can be calculated using
\begin{equation} \label{eqn:lrad}
\frac{dL_{rad}}{dt} = EM \times \Lambda(T) \hspace{2.5mm} \mbox{erg s$^{-1}$,}
\label{eqn:rad_loss_rate}
\end{equation}
where $\Lambda(T)$ is the radiative loss function. Here, this was estimated using tables of radiative loss rate as a function of emission measure for various temperatures, which were generated using CHIANTI \citep[v6.0.1][]{dere09} and the methods of \citet{cox69}.  This technique may lead to under- or over-estimates of the true radiative loss rate because it assumes an isothermal plasma which may not well approximate the flare's differential emission measure (DEM) distribution.  

In the above calculations, coronal abundances \citep{feld92} and the ionization equilibria of \cite{mazz98} were assumed. \citet{dere09} justified the use of these equilibria by comparing them to others obtained from the ionization rates of \citet{dere07} and a revised set of recombination rates.  The results were found to be similar.  A constant density of 10$^{10}$~cm$^{-3}$ was also assumed justified by \citet{whit05} who used CHIANTI to compute the spectrum of an isothermal plasma at 10~MK with densities of $10^9$, $10^{10}$, and $10^{11}$ cm$^{-3}$, and found no significant differences between them.

Finally, having calculated the radiative loss rate, the total radiative losses of the X-ray emitting plasma in the flare can be calculated by integrating between the flare start and end times:
\begin{equation} \label{eqn:lradint(t)}
L_{rad} = \int_{t_s}^{t_e} \frac{dL_{rad}(t)}{dt} dt \hspace{2.5mm} \mbox{ergs}
\end{equation}

\section{Background Subtraction Method}
\label{sec:method}
As the \goes lightcurves do not include any spatial information, they contain contributions not only from the flare but also from all non-flaring plasma across the solar disk.  In addition, the lightcurves include non-solar contributions such as instrumental affects which vary between the individual X-Ray Sensors.  These various background contributions can cause significant artifacts when deriving flare properties.  Therefore, it is imperative to isolate the flare signal from these contributions, particularly for weaker events.  In this study a generic temperature and emission measure-based background subtraction method (TEBBS) has been developed which improves upon the methods of \cite{born90}. These methods aim to discard possible flare signals which do not preserve the increasing nature of temperature and emission measure during the flare's rise phase (highlighted in Section~\ref{sec:intro}).  In this section, the limitations of previous background subtraction methods are discussed, before the TEBBS method is described in detail.

\begin{figure}
\begin{center}
\includegraphics[width=8cm]{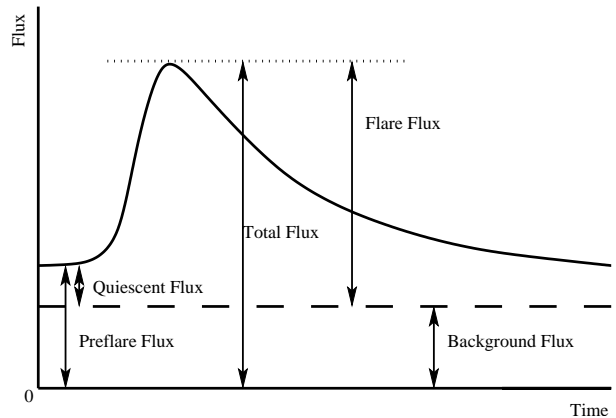}
\caption{Schematic of a flare X-ray lightcurve showing how the total flux detected by, e.g., the \goes XRS, is divided into constituent components. (Adapted from \citealt{born90}).  The total flux (solid line) is the sum of the flux from the flare plus the solar background (divided by the dashed line). The pre-flare flux, however, is the sum of the background component and the quiescent component of the flaring plasma (e.g., the associated active region).}
\label{fig:Born}
\end{center}
\end{figure}

\begin{figure*}
\begin{center}
\includegraphics[angle=90,width=14cm]{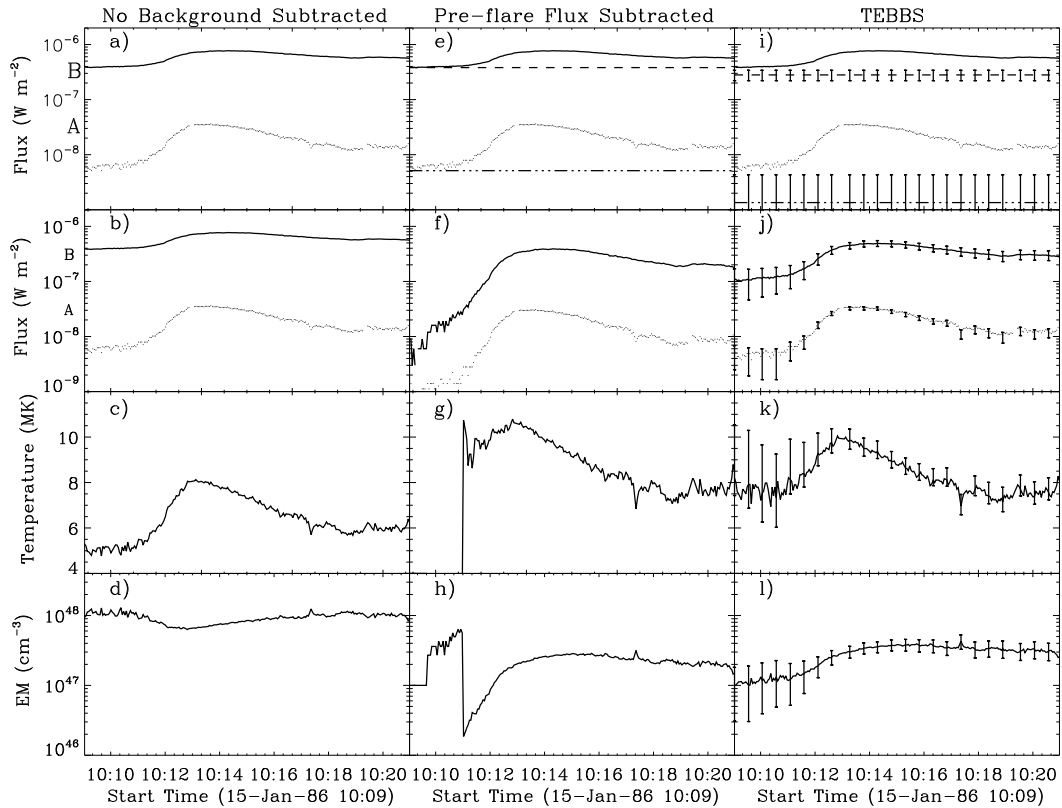}
\caption{\goes lightcurves and associated temperature and emission measure profiles for a B7 flare which occurred on 1986 January 15. The profiles in Figures~\ref{fig:threemethod}{\it a}--\ref{fig:threemethod}{\it d} are not background subtracted. The profiles in Figures~\ref{fig:threemethod}{\it e}--\ref{fig:threemethod}{\it h} have had the pre-flare flux in each channel subtracted, while Figures~\ref{fig:threemethod}{\it i}--\ref{fig:threemethod}{\it l} show the profiles obtained using the TEBBS method.  The error bars represent the uncertainty quantified via the range of background subtractions found acceptable by TEBBS.}
\label{fig:threemethod}
\end{center}
\end{figure*}

\subsection{Previous Background Subtraction Methods}
\label{sec:prevmeths}
The schematic in Figure~\ref{fig:Born} shows how a hypothetical \goes lightcurve is divided into its flaring and non-flaring (i.e., background) components.  The two limiting cases in calculating the the boundary between background and flare fluxes are either to assume that the total flux is dominated by the flare, thereby not performing any background subtraction, or to assume that the background is equal to the flux near the beginning of the event (`pre-flare' flux). The first assumption may be valid for events which are orders of magnitude above the background level, but is clearly incorrect for weaker events. The second assumption may be incorrect as there may be significant flare emission before the flare detection algorithm reports the start time. An example of the first method can be found in \cite{feld96b} in which no background was subtracted. An example similar to the second method can be used in the \goes workbench\footnote{\url{http://hesperia.gsfc.nasa.gov/rhessidatacenter/complementary\_data/goes.html}} which allows the background to be calculated as a line (polynomial or exponential) between the flux values at the start and end times of a flare.

\goes observations of a B7 flare which occurred 1986 January 15 at 10:09~UT, are shown in Figure~\ref{fig:threemethod}. In the first column, the flare signal is assumed to dominate, i.e.\ background is set to zero, while in the second column, the flare signal has been extracted by subtracting the pre-flare flux from the original lightcurve.  The top row (Figures~\ref{fig:threemethod}{\it a} and \ref{fig:threemethod}{\it e}) shows the non-background subtracted lightcurves with the background levels overplotted as horizontal lines.  The second row (Figures~\ref{fig:threemethod}{\it b} and \ref{fig:threemethod}{\it f}) shows the lightcurves after background subtraction. (N.B. since the background in the left column is zero, Figures~\ref{fig:threemethod}{\it a} and \ref{fig:threemethod}{\it b} are the same.) The third and fourth rows show the temperature and emission measure profiles, respectively, derived from the lightcurves shown in the second row.  An acceptable temperature profile is shown in Figure~\ref{fig:threemethod}{\it c} which peaks at 8~MK around 10:13~UT.  However the corresponding emission measure (Figure~\ref{fig:threemethod}{\it d}) {\it decreases} at the time of the flare. The reason for this uncharacteristic behavior is that the background flux component is dominating the emission measure evolution of the flare, thereby making it impossible for properties to be derived accurately. Conversely, by subtracting the pre-flare flux, as shown in Figures~\ref{fig:threemethod}{\it e}--\ref{fig:threemethod}{\it h}, significant artifacts are introduced to both the temperature and emission measure profiles.  This is because this background subtraction is causing the flux ratio at the beginning of the flare to be comprised of two small numbers, which leads to large discontinuities when folded through the temperature and emission measure calculations.

A more accurate approach would be to assume that the flare flux may also contain some contribution from the quiescent plasma from which it originates as shown in Figure~\ref{fig:Born}. This assumption was the basis for the background subtraction method developed by \cite{born90}. This technique applies three tests to a given combination of long and short channel background values: the increasing temperature test, the increasing emission measure test (together known as the increasing property tests), and the hot flare test; to determine whether a given choice of background levels produces physically meaningful results. The increasing property tests assume that both temperature and emission measure exhibit a characteristic overall increase during the rise phase. In these tests, background levels were selected and a preliminary subtraction was made. The relationship between the long and short channel fluxes during the rise phase was approximated with a linear fit of the form, $F_S = mF_L + c$, where $m$ is the slope and $c$ is the intercept.  From these fitted values, the temperature and emission measure for each point along the rise phase were were calculated using the polynomials of \citet{thom85} and compared to their previous value. If overall increases in these parameters were observed, then the background subtraction was said to have passed the increasing property tests.

To pass the hot flare test of \cite{born90}, the background temperature (calculated by plugging the ratio of the background values, $R_B = F^B_S/F^B_L$, into the temperature polynomial of \citealt{thom85}) must be less than the background subtracted flare temperature at all times during the flare. This helps prevent unphysical temperatures/emission measures being derived if the short channel approaches the detection threshold.

The tests of \cite{born90} were the first attempt to isolate a \goes flare signal from the background contributions based on the validity of the results produced. However, they have their drawbacks. The tests use the simple parameterizations of \cite{thom85} to calculate temperature and emission measure. Since then, \cite{whit05} devised tables from updated detector responses from \goess-1 to \goess-12 and plasma source functions that take into account the marked differences in temperature and emission measure when derived using coronal rather than photospheric abundances. Previously, \cite{thom85} had only provided one set of coefficients to their parameterization.  In addition, Bornmann's tests do not take into account the \goes instrumental temperature threshold. This threshold stands at 4~MK and exists because such a temperature would correspond to a flux ratio of $R=1/100$ which is beyond the sensitivity of the XRSs used in this study.  This means that these tests may not always identify the background combinations which may lead to unphysical profiles. Another shortcoming lies in the linear fit to the rise phase used in the increasing property tests. When demonstrating the method, \cite{born90} did not include the beginning of the rise phase in the linear fit because significant flux increases are often not observed there (e.g., Figure~\ref{fig:goodprofiles}{\it a}) and can affect the fit's accuracy. However, this leaves the beginning of the rise phase untested, which is the period most likely to exhibit spikes or discontinuities due to an unsuitable background subtraction. If these spikes are big enough, they can easily be mistaken for the true peaks and produce unreliable results. In the next section the TEBBS method is described in detail and the ways in which it improves upon the above-mentioned short-comings of the Bornmann tests are discussed.

\subsection{Temperature and Emission measure-Based Background Subtraction (TEBBS)}
\label{sec:tebbs}
TEBBS has been developed to facilitate accurate calculation of the plasma properties of large numbers of solar flares observed by \goess.  This is done by automatically isolating the flare signals from their background contributions.  Bornmann's method has been updated and improved in a number of ways. Firstly, explicit temperature and emission measure values calculated using \cite{whit05} are utilized in the background tests.  This is favored over simply using the flux ratios because the characteristic temperature and emission measure evolution of the flare can be directly analyzed. Secondly, extra criteria have been added to the hot flare test so that the minimum background subtracted flare temperature must be greater than the instrumental temperature threshold of 4~MK. Similarly the maximum background subtracted temperature must be less than the upper limit of the \citet{whit05} tables, i.e., $<$100~MK (corresponding to a flux ratio $<$1).  This upper limit is much higher than any \goes temperatures found by previous studies. This helps to identify all possible flare signals which produce unphysical profiles including those with discontinuities. Finally, another criterion has been added to the increasing property tests requiring that any temperature/emission measure value taken from the early rise phase which is not used in the linear fit, must be less than the peak taken from the rest of the rise phase. This helps to remove possible flare signals which show spikes at the beginning of the temperature/emission measure profiles which may not be identified by the original Bornmann tests. Both the TEBBS method and the original Bornmann tests assume that the background level in each channel is constant during the flare. This may not necessarily be the case, especially when the flare occurs during the decay of an earlier event.  This was deemed to be a rare enough occurrence (236 out of 34,361 events between 1991 and 2007, i.e., 0.7\%~--~see~Section~\ref{sec:gev}) that it would not introduce any significant errors. Moreover, as the peak flux and peak temperature occur near the beginning of an event, the slope of the background would have a negligible effect.

\begin{figure}
\begin{center}
\includegraphics[width=9cm]{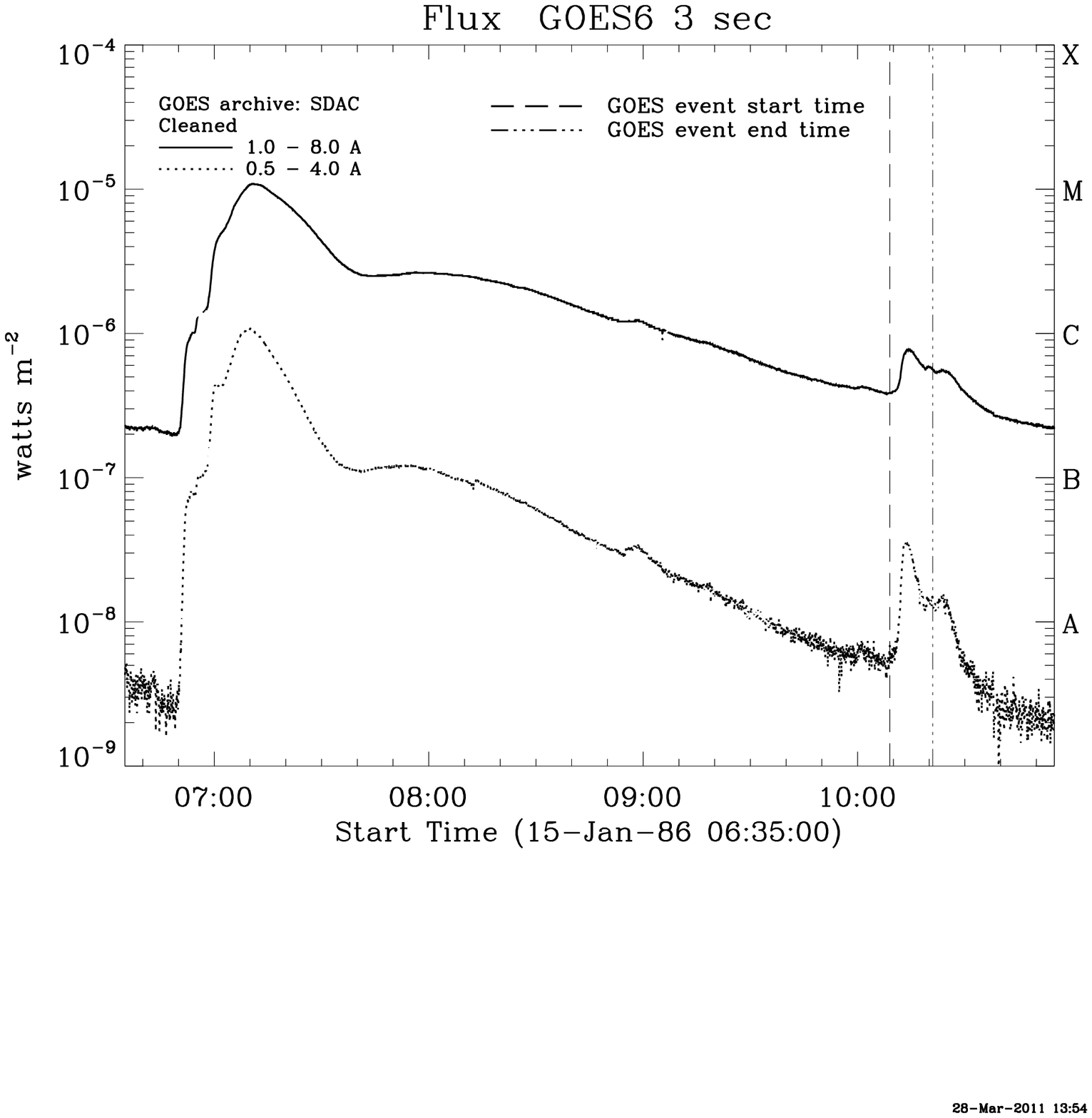}
\caption{GOES XRS lightcurves from 1986 January 15 06:35--10:55~UT. The start and end times of the B7 flare shown in Figures~\ref{fig:threemethod} and \ref{fig:svsl} as defined by the \goes event list are marked by the dashed and dot-dashed vertical lines respectively.}
\label{fig:beforeflare}
\end{center}
\end{figure}

The assumption underlying TEBBS is that the boundary between flare flux and background lies somewhere between zero and the pre-flare flux.  \cite{born90} justified this by assuming a quiescent flux component from the flare plasma.   However, there are a number of reasons why the background may not be well represented by the pre-flare flux.  For example, if the recorded start time of the flare is later than the true start time, the flare flux will have already risen considerably, thereby causing the pre-flare flux to be much higher than the actual background. Furthermore, if the flare occurs on the decay phase of another flare or at a time of high background flux, the flare will not be seen in the XRS data until the flare flux dominates the flux from the rest of the solar disk. Thus the flux at the reported flare start time (i.e., pre-flare flux) is a convolution of background and early flare flux, and therefore should not be subtracted in its entirety.  This was the case for the B7 flare on 1986 January 15. \goes lightcurves from an extended period around the flare (06:45--10:55~UT) are shown in Figure~\ref{fig:beforeflare}. The start and end times as defined by the \goes event list of the B7 flare are shown as the vertical dashed and dot-dashed vertical lines respectively. It can be seen that this flare has occurred on the decay phase of an M-class flare which began around 06:50~UT. Because of this high pre-flare flux, the initial evolution of the B7 flare was not readily detectable in the XRS data. Therefore the pre-flare flux was not an accurate approximation of the background and thus only a certain fraction should be subtracted.

\begin{figure}
\begin{center}
\includegraphics[angle=90,width=8cm]{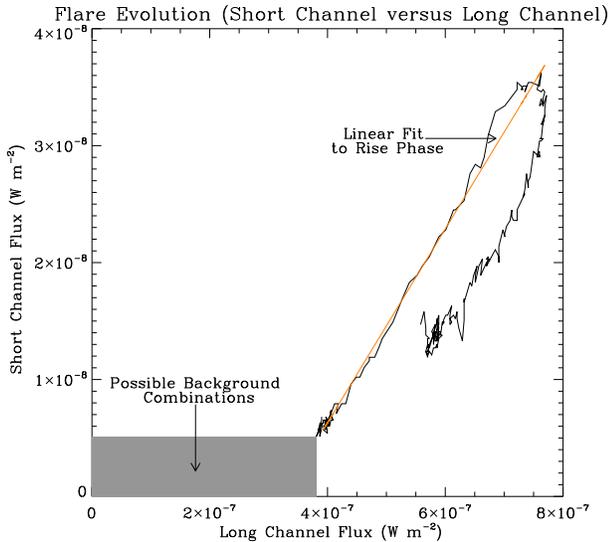}
\caption{Short channel flux versus the long channel flux for the 1986 January 15 flare (solid curve). The grey shaded area in the bottom left hand corner represents the possible combinations of background values from each channel for this event. The orange line represents a linear least-squares fit to the rise phase of the event.}
\label{fig:svsl}
\end{center}
\end{figure}

\begin{figure}
\begin{center}
\includegraphics[width=8cm]{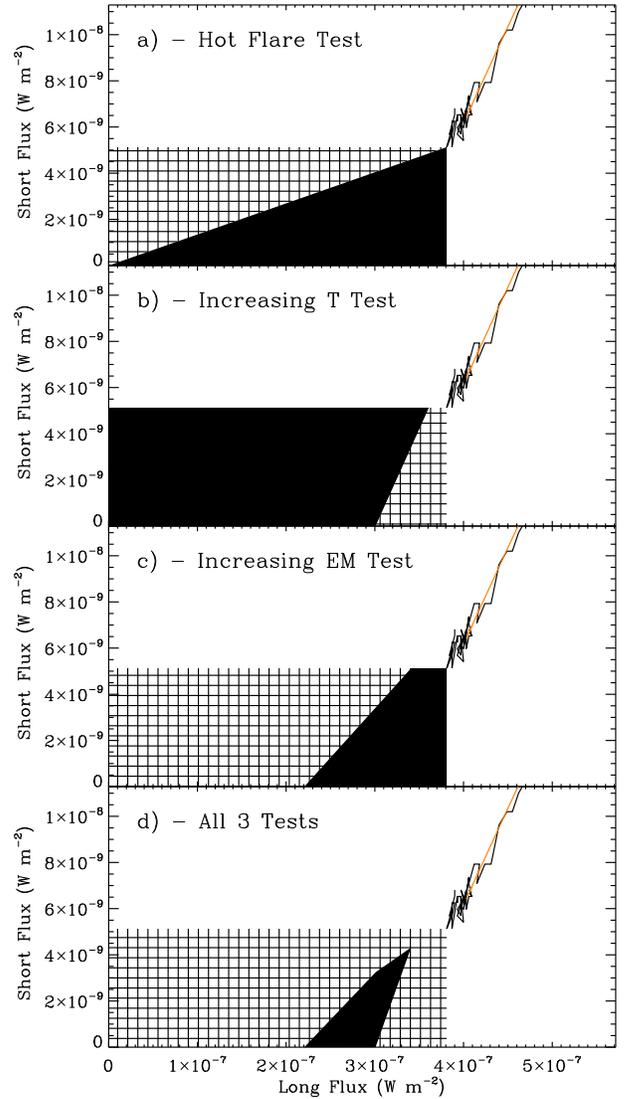}
\caption{Sample background space for 1986 January 15 flare.  The black shaded areas illustrate the range of values which pass a given background test, while the hashed regions denote background values which fail. a) the hot flare test, b) the increasing temperature test; c) the increasing emission measure test, and d) points which passed all three, or failed one or more.}
\label{fig:threetest}
\end{center}
\end{figure}

\begin{figure*}
\begin{center}
\includegraphics[angle=90,width=12cm]{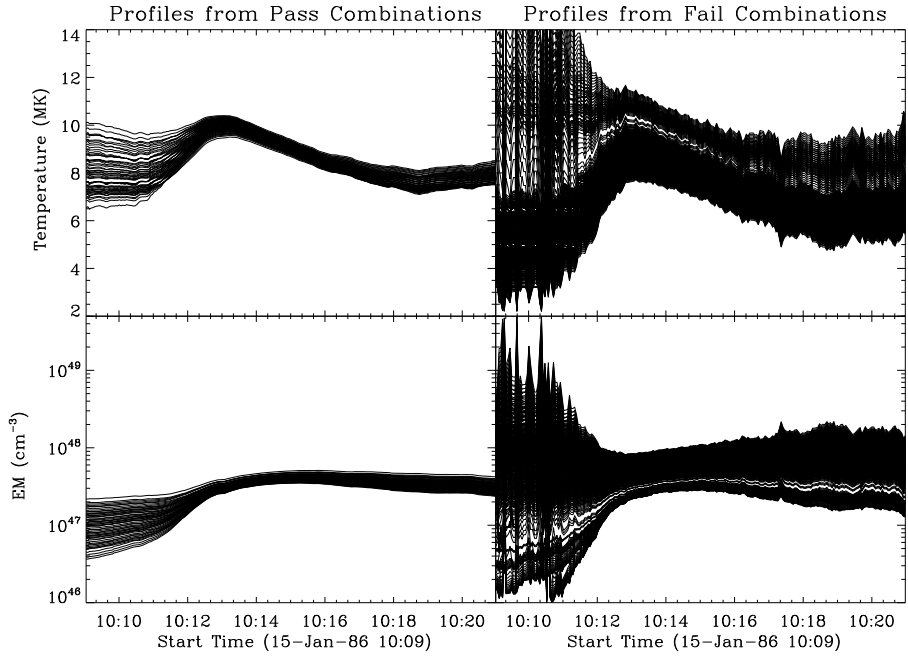}
\caption{Temperature and emission measure profiles for the 1986 January 15 flare for all possible background combinations.  The left column shows profiles which passed all three tests, while the right column shows profiles which failed one or more tests.}
\label{fig:profdists}
\end{center}
\end{figure*}

In order to apply the TEBBS method to this or any other flare, the first step is to define a sample space of possible background combinations. The range in each channel is between zero (equivalent to no background subtraction) and the minimum flux measured during the flare (equivalent to subtracting the pre-flare flux). The grey region in the bottom left of Figure~\ref{fig:svsl} shows the sample background space for the 1986 January 15 flare. This sample space is divided into twenty equally linearly separated discrete values in each channel $(F_L^B, F_S^B)$, thereby creating four hundred possible background combinations. Results were found to be independent of this binning and so twenty was chosen minimize computational time while ensuring that the background space was adequately sampled. Each background combination is then subtracted, thus creating four hundred sets of background subtracted lightcurves as possibilities for the flare signal. It is to these lightcurves that the hot flare test and increasing property tests are applied.

The first test to be applied is the hot flare test. The minimum temperature, $T_{min}$, of each lightcurve is calculated. Any background combinations corresponding to temperature profiles with a minimum temperature of $T_{min} \leq$~4~MK are discarded. Then $T_{min}$ is compared with the background temperature, $T_B$, calculated using the background values, $(F_L^B, F_S^B)$. If $T_{min} \leq T_B$ then that background combination is discarded. Furthermore, should the flux ratio at any point be greater than or equal to unity (i.e., $T \geq 100$~MK) the background combination is also discarded.  The background combinations of the 1986 January 15 flare which passed (solid region) and failed (hashed region) the hot flare test are shown in Figure~\ref{fig:threetest}a.  From this panel it can be seen that the number of possible background combinations has already been halved.

Next, the increasing property tests are applied. As in the Bornmann tests, the relationship between the short and long channel fluxes during the rise phase is fitted with linear function of the form, $F_S=mF_L+c$, so as to reduce the influences of fluctuations in the data.  Such a fit is justified by the fact that 90\% of flares in this study have a Pearson correlation coefficient greater than 0.85 for their rise phases and 95\% of flares have a correlation coefficient greater than 0.75.  Following \citet{born90}, the first sixth of the rise phase duration was not included in the linear fit (orange line, Figure~\ref{fig:svsl}). This is because significant increases are often not observed directly after the \goes event list start time (see Figure~\ref{fig:goodprofiles}) which can affect the accuracy of the fit to the rise phase.  The choice not to include the first sixth was determined by experiment in order to exclude any non-increasing portion of the early rise phase. Using these linearly approximated values, the evolution of the temperature and emission measure during the rise phase is calculated from all of the background subtracted lightcurves. Each value is compared with its preceding one and the percentage of times when the temperature/emission measure increases is calculated. Background combinations which result in profiles exhibiting a total increase less than a certain threshold are discarded. This threshold was chosen heuristically to be the maximum rise percentage from all four hundred background combinations minus seven e.g.\ if the maximum number of times that an increase in temperature was observed as a percentage of the total rise phase is 77\%, the threshold would be 70\%.  If this threshold leaves no background combinations which pass all three background tests it is iteratively reduced in steps of five percent until there is at least one background combination which passes all three tests.  This method still leaves the very beginning of the rise phase untested.  Therefore the next step is to calculate the temperature and emission measure profiles for the whole rise phase from the non-fitted background subtracted lightcurves. Any profiles that show a peak in the `untested' section of the rise phase greater than the peak found in the `tested' section are discarded. Figure~\ref{fig:threetest}{\it b} and \ref{fig:threetest}{\it c} show the background combinations which passed (solid regions) and failed (hashed regions) the increasing temperature and increasing emission measure tests respectively.

Having completed these tests, only background combinations which pass all three are deemed suitable. This leaves a small distribution of allowed background combinations, shown as the solid region in Figure~\ref{fig:threetest}d. The temperature and emission measure profiles corresponding to each of these background combinations are shown in Figure~\ref{fig:profdists} (smoothed for illustrative purposes). The left column shows the profiles corresponding to background combinations which passed all three tests, while the right column shows profiles corresponding to combinations which failed one or more tests. It can be seen that all the profiles in the left column are well behaved and are more conducive to calculating peak values and peak times. In contrast, many of the profiles in the right column exhibit discontinuities and spikes, particularly near the beginning of the flare. It is impossible to calculate useful peak values and times from these profiles, either automatically, or even manually. Although some of the temperature profiles in the right column appear to be well-behaved, the corresponding emission measure profiles contain artifacts, and vice versa.

The TEBBS method was applied to the 1986 January 15 B7 flare and the resulting time profiles are shown in the third column of Figure~\ref{fig:threemethod} (\ref{fig:threemethod}{\it i}--\ref{fig:threemethod}{\it l}).  The background levels were chosen from the combination closest to the center of the pass distribution in Figure~\ref{fig:threetest}{\it d} and are shown as the horizontal dashed and dot-dashed lines in Figure~\ref{fig:threemethod}{\it i}.  The background subtracted fluxes can be seen in Figure~\ref{fig:threemethod}{\it j}.  The error bars mark the uncertainty in the background subtraction which was taken as the range of the pass distribution in Figure~\ref{fig:threetest}{\it d}.  The TEBBS temperature and emission measure profiles are shown in Figures~\ref{fig:threemethod}{\it k} and \ref{fig:threemethod}{\it l}, respectively. Both of these profiles show smooth rise and decay phases. Note that the temperature profile does not have a discontinuity as in Figure~\ref{fig:threemethod}{\it g}. Furthermore, the emission measure evolution is no longer dominated by the background contribution as in Figure~\ref{fig:threemethod}{\it d} nor does it exhibit spikes or discontinuities as in Figure~\ref{fig:threemethod}{\it h}.  The error bars in these panels represent the range of acceptable temperature and emission measure profiles seen in the left column of Figure~\ref{fig:profdists}.  Note that the uncertainties at the beginning of the flare are largest.  This is expected as the flux ratio during the early rise phase is made of smaller numbers than the rest of the flare.  Therefore, a slight inaccuracy in the background subtraction can cause a more significant change in temperature and emission measure.  The beginning of the flare also shows an unexpectedly high temperature of $\gtrsim$6~MK.  This can be explained by the fact that the start time defined by the \goes event list was probably after the actual start time due to the high emission from the M-class flare which preceded it (see Figure~\ref{fig:beforeflare}). This would imply that the flaring plasma was initially cooler than 6~MK but by the time the flare emission could be detected over that of the M-class flare, the plasma had already been heated substantially.  Figure~\ref{fig:threemethod} shows that TEBBS has performed a successful, automatic background subtraction, superior to either of those performed with the other two methods discussed in Section~\ref{sec:prevmeths}.

Having successfully tested TEBBS on other flares chosen at random, the method was applied to all 51,196 selected flares in the \goes event catalog from 1980 January 1 to 2007 December 31. The specific background combinations were chosen in the same way as for the 1986 January 15 flare.  Of these, successful background subtractions could not be performed for 1,870 events ($\sim$3\%) and so were discarded from the dataset. Of these 144 were `double flares' or even `triple flares' making the assumptions of TEBBS invalid.  The remaining 1,726 can be characterised in the following ways: events dominated by `bad' points; events where the short channel approaches the lower detection threshold leading to unphysical flux ratios and hence derived properties; unphysical light curves resembling square or triangular waves with up to an order of magnitude flux amplitude; events which we would be interested to analyze.  The size distribution of events discarded by TEBBS was very similar to that of the original dataset. This shows that TEBBS did not preferentially discard flares of a particular size and therefore did not bias the results of this study. The associated plasma properties (peak temperature, peak emission measure, radiative loss rates, and total radiative losses) were derived for the remaining 50,703 events.  Uncertainties on the plasma properties for each event were calculated from the corresponding  range of allowed TEBBS background subtractions as was done in Figures~\ref{fig:threemethod}{\it j}--\ref{fig:threemethod}{\it l}.  Values from `bad' data points and neighboring data points were ignored due to their tendency to produce unreliable spikes when folded through the temperature and emission measure calculations.  `Bad' points are marked as such by the \goes software as untrusted measurements because of the instrument states reported by telemetry (e.g., because of gain changes) or because they are outliers from surrounding data.  The identification of these `bad' points is justified by simultaneous observations from other \goes spacecraft.  The TEBBS database is publicly available on Solar Monitor\footnote{\url{http://www.solarmonitor.org/TEBBS/}} and will also be available through the Heliophysics Integrated Observatory\footnote{\url{http://www.helio-vo.eu/}}.
The statistical relationships between the above derived properties are discussed in the next section.

\section{Results}
\label{sec:results}
\begin{figure*}
\begin{center}
\includegraphics[angle=90,width=15.5cm]{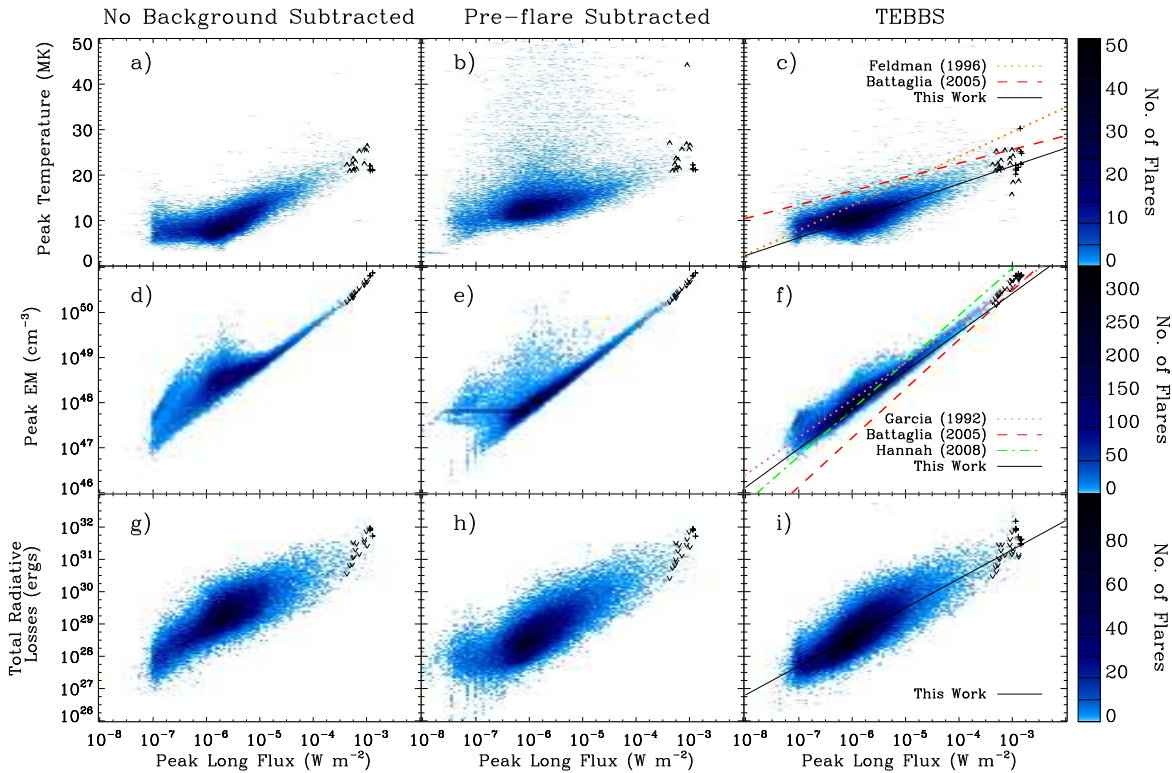}
\caption{Two-dimensional histograms of peak temperature, peak emission measure, and total radiative losses, each as a function of peak long channel flux, derived using various background subtraction techniques for all selected \goes events observed between 1980 and 2007. The data in the left column has not had the background subtracted. The data in the middle column had the pre-flare flux subtracted, while the distributions in the right column have been derived using the TEBBS method. Overplotted on panels {\it c} and {\it f} are the relationships derived by \citet[purple dotted line]{garc92}, \citet[orange dotted line]{feld96b}, \citet[red dashed line]{batt05}, and \citet[green dot-dashed line]{hann08}. The solid black lines show relationships described by Equations~\ref{eqn:t(f)fit}, \ref{eqn:em(f)fit} and \ref{eqn:lradfit}.  The arrow heads mark events which are upper or lower limits due to saturation of the XRS and point in the direction that the true value would have been located. The crosses mark events which display lower flux limits and derived properties which are neither upper nor lower limits but only rough estimates. This is because their derived properties are functions of the fluxes in both channels which each saturated.  See Appendix for a more detailed discussion of the effects of saturation on derived flare plasma properties and the saturation levels of the XRS on the various \goes satellites.}
\label{fig:plasma}
\end{center}
\end{figure*}

Peak temperature, peak emission measure, and total radiative losses each as a function of peak long channel flux are shown as a density of points in Figure~\ref{fig:plasma}.  Each column shows distributions obtained using each of the three background subtraction methods discussed in Section~\ref{sec:method}.  Uncertainties on each data point are not shown for clarity.

The relationship between peak flare temperature and peak long channel flux is shown in Figures~\ref{fig:plasma}{\it a}--\ref{fig:plasma}{\it c}. While the non-background subtracted distribution in Figure~\ref{fig:plasma}{\it a} displays some trend of larger flares exhibiting higher temperatures, there is a flattening of the distribution below the C1 level. This is due to the influence of the background contributions which become highly significant at low fluxes. In addition, there is a stark vertical edge at B1 level due to the absence of A-class flares in the \goes event list.  A horizontal edge at $\sim$5~MK is also seen, which is due to the instrumental detection limit. There is more scatter in the pre-flare background subtracted distribution in Figure~\ref{fig:plasma}{\it b}, with events of all classes showing temperatures in excess of 25~MK up to a temperature of $\sim$80~MK (beyond the range of the plot axis). By subtracting all of the pre-flare flux, the value of the flux ratio at the beginning of the flare can become erroneously large due to dividing one small number by another. This can lead to spuriously high temperature values when folded through the temperature calculations and can often be greater than the real peak temperature (see Section~\ref{sec:method}). Many of the high temperature values ($>$25~MK) in Figure~\ref{fig:plasma}{\it b}, particularly those corresponding to low peak long channel fluxes, have been taken from such spikes early in the flare. In contrast, the TEBBS method performs background subtractions which tend not to cause such temperature spikes. As a result the distribution in Figure~\ref{fig:plasma}{\it c} shows much less scatter. The flattening at low fluxes seen in Figure~\ref{fig:plasma}{\it a} has also been reduced by the use of the TEBBS method.

In order to examine the relationship between peak temperature and peak long channel flux, a number of methods were used.  First, the Kendall tau coefficient was calculated which is a non-parametric correlation coefficient i.e.\ it does not assume a pre-defined model for the data.  It is based on the rank of the data points rather than the values themselves, making it more suitable than other correlation coefficients (e.g.\ the Pearson linear coefficient) to distributions with significant outliers or scatter such as those in this study.  The Kendal tau correlation coefficient for Figure~\ref{fig:plasma}{\it c} was found to be 0.42 which represents a statistically significant correlation.  Next, the relationship was quantified using linear regression.  Ordinary least squares (OLS) was not used however, because of three characteristics of the data: the presence of several outliers which produced non-normal behavior between an OLS regression fit and the observations (i.e.\ the residuals were not normally distributed); data truncations due observational cutoffs below B1 level and 4~MK; the underlying power-law number distribution of the observations, i.e.\ the greater number of smaller events relative to larger ones.  To address these characteristics of the observations, the methods of robust statistics were used. The basic assumption in OLS is that the residuals are normally distributed and the solution to the problem is calculated by minimizing the sum of the squared residuals. However in this case, the sum is replaced by the median of the squared residuals. This results in an estimator that is resistant to the outliers by finding the narrowest strip covering half the observations \citep{rous84}. To account for the population distribution, the regression analysis was weighted using the flux values themselves, with smaller events weighted less than larger events. The truncation in the data is handled using the method of \citet{bhat83}. 
The form of the fit resulting from this method is given by
\begin{equation}\label{eqn:t(f)fit}
T = \alpha + \beta log_{10}F_L \mbox{    MK}
\end{equation}
This form was chosen because it implies a linear relationship between temperature and the $log_{10}$ of the peak long channel flux such as those found by both \citet{feld96b} and \citet{batt05}.  The values of $\alpha$ and $\beta$ were found to be 34$\pm$3 and 3.9$\pm$0.5 respectively.  Finally, the goodness of this fit was examined by using a modified, robust R$^2$ statistic which quantifies the variance in the data explained by the model.  Whereas the usual R$^2$ value is based on the mean-squared-error, the modified robust R$^2$ statistic is based on the median (consistent with the robust fitting method used above).  It also accounts for the degrees of freedom and in the fitting and the uncertainties on each data point.  It was found that the modified robust R$^2$ value for the above fit was 0.62.  This value is lowered by the structure in the distribution at least in part caused by the instrumental truncations below B-class and 4~MK.  Nonetheless this value still implies that the Equation~\ref{eqn:t(f)fit} is a suitable fit to the distribution.

\begin{figure*}[!t]
\begin{center}
\includegraphics[angle=90,width=16cm]{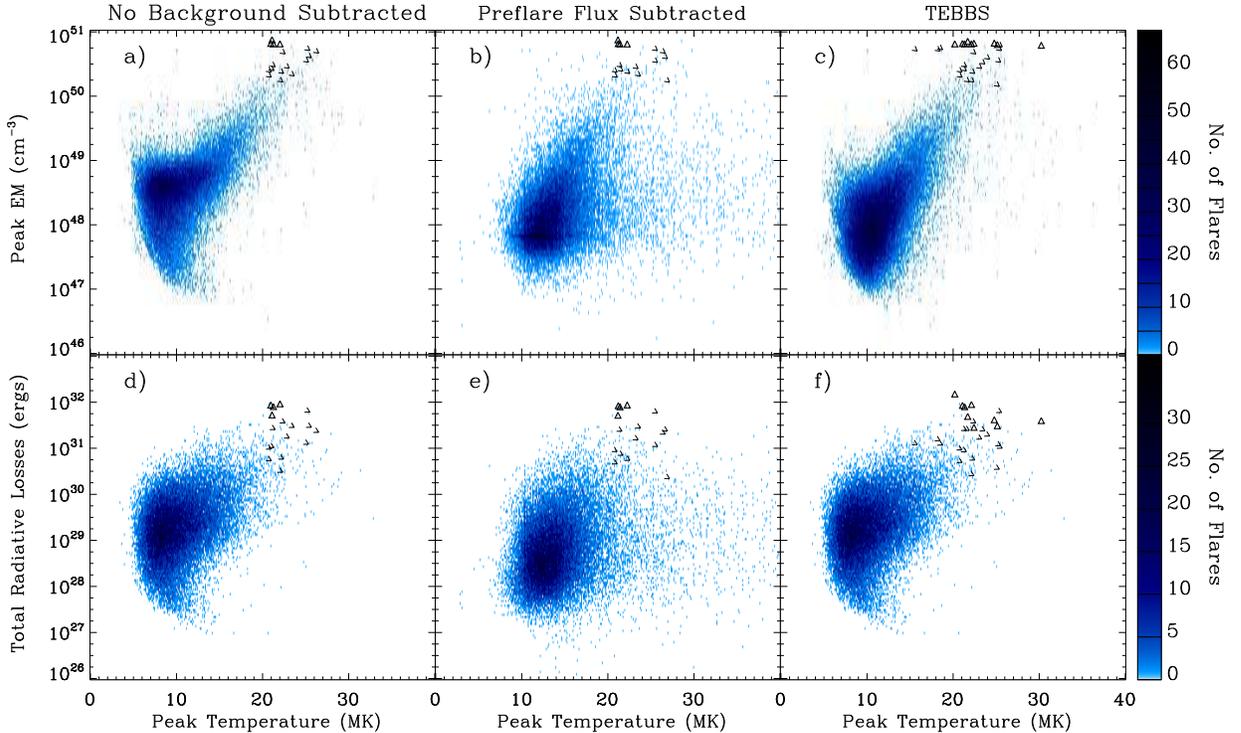}
\caption{Two-dimensional histograms of peak emission measure and total radiative losses as a function of peak temperature for each of the three background subtraction techniques.  The arrow heads mark events which are upper or lower limits due to saturation of the XRS and point in the direction that the true value would have been located. The triangles mark events whose values are only rough estimates due to both channels saturating.  See Appendix for a more detailed discussion of the effects of saturation on derived flare plasma properties and the saturation levels of the XRS on the various \goes satellites.}
\label{fig:t_vs_em}
\end{center}
\end{figure*}

The relationship between peak emission measure and peak long channel flux is shown as a density of points in Figures~\ref{fig:plasma}{\it d}--\ref{fig:plasma}{\it f}. The non-background subtracted distribution in Figure~\ref{fig:plasma}{\it d} displays the same selection effect as that in Figures~\ref{fig:plasma}{\it a}, i.e., a sharp cut-off at low \goes class ($\sim$B1 level). This cut-off is not seen as clearly in Figures~\ref{fig:plasma}{\it e} or \ref{fig:plasma}{\it f}) due to background subtraction.  Large amounts of scatter were found below the M1 level in Figures~\ref{fig:plasma}{\it d} and \ref{fig:plasma}{\it e} which is not seen to the same degree in the TEBBS distribution in Figure~\ref{fig:plasma}{\it f}. The unusually high emission measures in Figures~\ref{fig:plasma}{\it d} and \ref{fig:plasma}{\it e} have been recorded from erroneous features such as those in Figures~\ref{fig:threemethod}{\it d} and \ref{fig:threemethod}{\it h}. There is a well defined linear lower edge in all three of the distributions. A similar feature was found by \cite{garc88} and \cite{garc92}. This edge is a natural consequence of the way emission measure is calculated, approximated by Equation~\ref{eqn:em}.  The second term in this equation asymptotically tends to zero which means it varies very little at high temperatures.  As a result, emission measure becomes directly proportional to long channel flux causing the observed linear edge, which corresponds to high temperatures.  This feature was also seen by \cite{garc92} but not explained.

To examine the correlation between peak emission measure and peak long channel flux, the Kendall tau coefficient of the TEBBS distribution in Figure~\ref{fig:plasma}{\it f} was calculated and found to be 0.8, implying a significant correlation.  In order to compare our results with those of previous studies, a linear relationship between the $log_{10}$ of these two properties, such as those found by \citet{garc92}, \citet{batt05}, and \citet{hann08}, was applied to the data.  The fit was performed in log-log space using the same linear regression method used for the relationship between the temperature peak and the long channel flux.  To remain consistent with previous studies this fit was re-expressed as a power-law of the form:
\begin{equation}\label{eqn:em(f)fit}
EM = 10^{\gamma} F_L^{\delta} \mbox{   cm$^{-3}$}
\end{equation}
The values of $\gamma$ and $\delta$ were found to be 53$\pm$0.1 and 0.86$\pm$0.02 respectively.  This relationship is expressed in the inverse as
\begin{equation}\label{eqn:f(em)fit}
F_L = \eta EM^{\zeta} \mbox{   W m$^{-2}$}
\end{equation}
where $\eta$ and $\zeta$ were found to be 1$\times$10$^{-61\pm1}$ and 1.15$\pm$0.02 respectively.  The modified robust R$^2$ value for the above model was found to be 0.73, implying a good fit.

Total radiative losses as a function of peak long channel flux is shown as a density of points in Figures~\ref{fig:plasma}{\it g}--\ref{fig:plasma}{\it i}.  All three distributions clearly show an increasing trend with peak long channel flux.  The similarity between the three distributions suggests that total radiative losses are not as sensitive to background subtraction as either peak temperature or peak emission measure.  This is to be expected since peak values are taken from single points which can be very sensitive to erroneous spikes caused by inappropriate treatment of the background.  However, total radiative losses are integrated over the flare duration.  Therefore, if a flare contains erroneous temperature or emission measure spikes, their contribution to the total radiative losses will not be as significant if the rest of the flare is `well-behaved'.  For small flares, however, the effect of these erroneous values would be expected to have a greater influence.  The `turn-up' at A- and B-class levels in Figure~\ref{fig:plasma}{\it h} is consistent with this.  This distribution was found to have a high Kendall tau correlation coefficient of 0.73.  It was then fit using the same method as used for the emission measure peak long channel flux relationship.  The resulting fit was expressed in the form:
\begin{equation}\label{eqn:lradfit}
L_{rad} = 10^{\varepsilon} F_L^{\theta} \mbox{   ergs}
\end{equation}
This form was chosen because it implies a linear relationship between the log$_{10}$ of the total radiative losses and long channel peak flux with an intercept of $\varepsilon$ and a slope of $\theta$.  This is justified by a high Pearson correlation coefficient, calculated in log-log space as 0.8.  The values for $\varepsilon$ and $\theta$ were found to be 34$\pm$ 0.4 and 0.9$\pm$0.07 respectively.  The modified robust R$^2$ statistic was found to be 0.71, implying that Equation~\ref{eqn:lradfit} well represents the distribution.

Distributions of peak emission measure and total radiative losses as a function of peak temperature are shown in Figure~\ref{fig:t_vs_em}. Each column corresponds to distributions obtained from the same background subtraction methods as in Figure~\ref{fig:plasma}.

Peak emission measure as a function of peak temperature is shown as a density of points in Figures~\ref{fig:t_vs_em}{\it a}--\ref{fig:t_vs_em}{\it c}. A clear relationship between the two properties is not apparent in the non-background subtracted distribution in Figure~\ref{fig:t_vs_em}{\it a}. A horizontal edge from 5-12~MK and just above 10$^{49}$~cm$^{-3}$ is exhibited with the majority of  flares located just below this edge. Any relationship between these two properties is even less clear in Figure~\ref{fig:t_vs_em}{\it b}. Very large scatter extends beyond the range of this plot to $\sim$80~MK. The artifacts introduced into both the temperature and emission measure profiles by each of the respective background subtraction methods (such as those in Figures~\ref{fig:threemethod}{\it g} and \ref{fig:threemethod}{\it h}) have exacerbated the scatter. A more discernible trend with less scatter is revealed by the use of TEBBS in Figure~\ref{fig:t_vs_em}{\it c}. This distribution clearly shows that flares with hotter peak temperatures have greater peak emission measures. However, there seems to be an edge to this distribution at low temperatures which may also be explained by a limit of Equation~\ref{eqn:em}.

Total radiative losses as function of peak temperature is displayed as a density of points in Figures~\ref{fig:t_vs_em}{\it d}--\ref{fig:t_vs_em}{\it f}. Although the TEBBS distribution in Figures~\ref{fig:t_vs_em}{\it f} appears comparable to the non-background subtracted distribution in Figure~\ref{fig:t_vs_em}{\it d}, it displays less scatter than seen in the pre-flare subtracted distribution in Figure~\ref{fig:t_vs_em}{\it e} which has data points extending beyond the range of the plot axis to $\sim$80~MK.  No clear trend between peak temperature and total radiative losses is discernible in any of Figures~\ref{fig:t_vs_em}{\it d}--\ref{fig:t_vs_em}{\it f}, although Figures~\ref{fig:t_vs_em}{\it d} and \ref{fig:t_vs_em}{\it f} do show a tendency for higher temperature flares to have greater total radiative losses.  This implies there is no strong relationship between these properties.  This is despite the fact that total radiative losses are function of both temperature and emission measure.  This absence of a trend may be due to the assumptions used in deriving these properties (e.g.\ constant density).

\section{Discussion}
\label{sec:discussion}
The TEBBS distributions in Figures~\ref{fig:plasma}~and~\ref{fig:t_vs_em} consistently show the least scatter and most discernible trends between properties derived from \goes data. The non-background subtracted and pre-flare subtracted distributions show a higher number of artifacts such as edges and anomalously high values. This shows that TEBBS is a superior method of automatically subtracting background than either of the other two methods; first because it successfully separates the flare signal from the background contributions, and second, produces fewer artifacts in doing so. However, there still may be biases in the distributions derived using TEBBS. Such biases may be due to the fact that TEBBS uses full-disk integrated observations. A comparison between temperature and emission measure profiles produced in this study and those derived from spatially resolved instruments could further highlight how reliable the TEBBS results are and be used to quantify any systematic biases. Spatially resolved observations could be taken from instruments which observe in similar wavelength bands to the XRS, such as the Soft X-ray Telescope (SXT) onboard {\it Yohkoh}, the Soft X-ray Imager (SXI) onboard \goess-12 and \goess-13, or the X-Ray Telescope (XRT) onboard {\it Hinode}. Such a study would be useful in further determining the strengths and weaknesses of the TEBBS method.  Furthermore, it must be acknowledged that several necessary assumptions were used in calculating the plasma properties, as there are any time these properties are derived using \goes observations.  In this study, coronal abundances \citep{feld92}, a constant density of 10$^{10}$~cm$^{-3}$, the ionization equilibrium from \cite{mazz98}, and an isothermal plasma were assumed. It has been shown that coronal iron abundances during flares can reach eight times the photospheric level \citep{feld04}, or higher, and  \cite{phil10} found coronal densities above 10$^{13}$~cm$^{-3}$ using high-temperature density sensitive ratios. Using either of these assumptions in the calculation of the flaring plasma properties could affect the results.  However, this was not done in this study so that these results would be more comparable with those of previous studies.

The distribution in Figure~\ref{fig:plasma}{\it c} was compared with the studies of \citet{feld96b} and \citet{batt05}.  These studies found a linear correlation between peak temperature and the log$_{10}$ of peak long channel flux.  In both papers, the relation was expressed in the form:
\begin{equation}\label{eqn:prevt}
F_L = 3.5 \times 10^{\beta T + \kappa} \mbox{   W m$^{-2}$}
\end{equation}
The values of $\beta$ and $\kappa$ from these studies can be found in Table~\ref{tab:prevt}.  Values from this study were calculated by rearranging Equation~\ref{eqn:t(f)fit} into the form of Equation~\ref{eqn:prevt} and are also included in Table~\ref{tab:prevt} for comparison.  The \citet{feld96b} and \citet{batt05} relations are also overplotted on Figure~\ref{fig:plasma}{\it c} as the orange dotted and red short dashed lines respectively.  The distribution of this study reveals predominantly lower temperatures for a given long channel flux than both previous studies.  There is closer agreement with \cite{feld96b} than \citet{batt05} for B-class events but beyond this, lower temperatures are obtained.  This can be explained by \cite{feld96b} using the BCS to calculate temperature.  In that paper, it was stated that temperatures obtained with the BCS agreed with those from \goes below 12~MK but above this point were higher on average by a factor of 1.4. To investigate this, the mean peak temperature of all flares in our sample of M-class or greater was computed and found to be 16~MK. This flux threshold was chosen for two reasons.  Firstly, the difference between background subtracted (this study) and non-background subtracted (\citealt{feld96b}) results are negligible in this regime. Secondly, of these events, 95\% had peak temperatures greater than 12~MK.  This was compared to the mean Feldman temperature obtained for these events by plugging their long channel peak flux into the fit of \citet{feld96b}.  The Feldman mean temperature was found to be 20.9~MK which differs from that of this study by a factor of 1.3.  This is lower than the value quoted by \cite{feld96b}.  This is because they measured temperature at the the time of the long channel peak which would be expected to occur after the temperature peak.  Assuming that the temperature peaks before the long channel flux, the difference in ratios would imply that a flare's temperature (M-class or greater) drops by 10\% before the long channel peak.  This implication is supported by this study, as part of which the temperature at the time of the long channel peak was also calculated.  This shows that between the temperature and long channel peaks, a flare's temperature drops on average by 10\%--11\% for flares greater than or equal to M-class and 9\%--10\% for all flares.

\begin{table}
\begin{center}
\caption[Values for $F_L$-$T$ relationship (Equation~\ref{eqn:prevt})]{Values for $F_L$-$T$ relationship (Equation~\ref{eqn:prevt})}
\label{tab:prevt}
\begin{tabular}{ccc}
\hline
\hline
Study		&$\beta$				&$\kappa$		\\
\hline
\cite{feld96b}	&0.185				&-9				\\
\cite{batt05}	&0.33$\pm$0.29		&-12				\\
TEBBS		&0.26$^{+0.04}_{-0.01}$	&-9$^{+1}_{-2}$	\\
\hline
\end{tabular}
\end{center}
\end{table}

The slope of the relation of \citet{batt05} appears closer to that of the fit found by this study.  However, the relation consistently gives temperatures which are 3--4~MK higher.  This discrepancy in the intercept is due to \cite{batt05}'s use of \rhessi to obtain the temperatures.  The value of $T$ in that study was calculated as either the temperature of an isothermal fit or the lower of two temperatures in a multi-thermal fit to a \rhessi spectrum.  This was compared to \goes temperature, $T_G$, and a relation was derived given by
\begin{equation}\label{eqn:t1tg}
T = 1.12 T_G + 3.12 \mbox{   MK}
\end{equation}
Substituting this into Equation~\ref{eqn:prevt} and rearranging into the form of Equation~\ref{eqn:t(f)fit}, values for $\alpha$ and $\beta$ are found to be 28$^{+198}_{-15}$ and 2.7$^{+17.3}_{-1.3}$ respectively.  Although these values are similar to those found for this study ($\alpha = 33 \pm 3$; $\beta = 3.9 \pm 0.5$), the large uncertainties mean that little statistical significance can be assigned to this similarity.  This highlights that a more comprehensive study than that of \cite{batt05} was needed to more precisely understand the statistical relationships between the thermal properties of solar flares.

Next, the emission measure distribution in Figure~\ref{fig:plasma}{\it f} was compared work of \cite{garc92}.  In that paper, the linear edge to this distribution was addressed (also discussed in Section~\ref{sec:results} of this paper) and a fit to this lower bound was quoted from a previous paper, \cite{garc88}.  This was of the same form as Equation~\ref{eqn:em(f)fit}.  Garcia's values are shown in Table~\ref{tab:emedge} and the fit corresponding to them is overplotted on Figure~\ref{fig:plasma}{\it f} as the purple dotted line.  However, this relation does not fit the lower bound of this distribution very well.  In fact it seems to better fit the distribution itself being as it is so similar to the values found for Equation~\ref{eqn:em(f)fit} in Section~\ref{sec:results} (shown in the second row of Table~\ref{tab:emedge}).  A rough fit to this edge shows it is much better formulated by the parameters shown in the third row of Table~\ref{tab:emedge}.  The discrepancy may be because the sample of \cite{garc92} was insufficient to reveal the actual limit of this relationship.  However, it may be also be due to the fact that methods different from those of \citet{whit05} were used to calculate temperature and emission measure \citep[e.g.][]{thom85}.  The credibility of this limit is important as it suggests a well defined minimum amount of material emitting in the \goes passbands produced by a flare of a certain long channel peak flux.  Although this limit is due to the nature of Equation~\ref{eqn:em} the result should be compared with statistical studies using other instruments to confirm whether it is a breakdown in the validity of the temperature and emission measure calculations of \citet{whit05} or has any physical significance.

\begin{table}
\begin{center}
\caption[Values for linear edge in $F_L$-$EM$ distribution (Same form as Equation~\ref{eqn:em(f)fit})]{Values for linear edge in $F_L$-$EM$ distribution (Same form as Equation~\ref{eqn:em(f)fit})}
\label{tab:emedge}
\begin{tabular}{ccc}
\hline
\hline
Study						&$\gamma$		&$\delta$			\\
\hline
\cite{garc92}					&53.04			&0.83			\\
TEBBS (Eqn~\ref{eqn:em(f)fit})		&53$\pm$0.1		&0.86$\pm$0.02	\\
TEBBS (lower edge)				&53.4			&0.96			\\
\hline
\end{tabular}
\end{center}
\end{table}

This distribution was also compared to the work of \cite{batt05} and \citet{hann08} who found correlations between \rhessi emission measure and background subtracted \goes long channel peak flux.  These relations were expressed in the same form as Equation~\ref{eqn:f(em)fit}.  The values found by these studies are displayed in Table~\ref{tab:prevem} along with the values from this study for comparison.  These previous fits are also overplotted on Figure~\ref{fig:plasma}{\it f} as the dashed red and dot-dashed green lines respectively.  These fits are steeper than our distribution.  The relation of \cite{hann08} however, agrees well at B- and C-class which was the range on which that study focused (A to low C-class).  The difference in slope can not be due to the different sensitivities of \goes and \rhessi as \cite{hann08} showed that \goes emission measure is consistently a factor of two greater than that obtained from \rhessis.  The difference in slope may therefore be attributed to the extension of the distribution to M- and X-class.  However, it may also have been affected by the fact that \citet{hann08} calculated the emission measure at the time of the peak in the \rhessi 6--12~keV passband rather than the peak emission measure, as in this study.  The relation of \cite{batt05} gives consistently lower emission measures than both the TEBBS distribution and relation of \cite{hann08}.  This can be explained by \cite{batt05} measuring the emission measure at the time of the hardest HXR peak which tends to occur early in the flare before the SXR and emission measure peaks.

\begin{table}
\begin{center}
\caption[Values for $EM$-$F_L$ relationship (Equation~\ref{eqn:f(em)fit})]{Values for $EM$-$F_L$ relationship (Equation~\ref{eqn:f(em)fit})}
\label{tab:prevem}
\begin{tabular}{ccc}
\hline
\hline
Study		&$\eta$							&$\zeta$			\\
\hline
\cite{batt05}	&3.6$\times$10$^{-50}$				&0.92$\pm$0.09	\\
\cite{hann08}	&1.15$\times$10$^{-52}$				&0.96			\\
TEBBS		&1$\times$10$^{-61\pm1}$			&1.15$\pm$0.02	\\
\hline
\end{tabular}
\end{center}
\end{table}

The distribution of peak emission measure as a function of peak temperature in Figure~\ref{fig:t_vs_em}{\it c} shows that hotter flares have larger peak emission measures.  \cite{feld96b} found a power-law relationship between these two properties expressed by
\begin{equation}\label{eqn:emtfeld}
EM = 1.7 \times 10^{0.13T + 46} \mbox{   cm$^{-3}$}
\end{equation}
However, the Pearson correlation coefficient between temperature and the log of emission measure in Figure~\ref{fig:t_vs_em}{\it c} was calculated to be 0.3.  This implies that these properties are in fact not well linearly correlated at all despite an apparent trend of hotter flares having higher emission measures.  \cite{hann08} also examined the relationship between emission measure and temperature for A--low C-class flares and found no correlation.  If the Figure~\ref{fig:t_vs_em}{\it c} distribution is examined more closely, there does not appear to be any relationship between the two properties within the range \cite{hann08} studied.  Indeed, the Pearson correlation coefficient for C1.0-class and below is only 0.2.  This supports \cite{hann08} findings.  However further examination of this relationship is necessary to draw firmer conclusions.

\section{Conclusions \& Future Work}
\label{sec:conclusions}
A method, TEBBS, has been presented for isolating the solar flare signal from \goes soft X-ray lightcurves by accounting for contributions from both solar and non-solar backgrounds.  This allows the properties of the flaring plasma itself to be more accurately derived. It can be systematically applied to any number of flares, removing many of the inconsistencies that can be introduced when manually defining a background level. This makes it a particularly suitable method for conducting large-scale statistical studies of solar flares characteristics. TEBBS was found to produce fewer spurious artifacts in the derived temperature and emission measure profiles for both individual events (Figure~\ref{fig:threemethod}) and in large statistical samples (Figures~\ref{fig:plasma}~and~\ref{fig:t_vs_em}), compared to when either all or none of the pre-flare flux was removed. This led to more reliable relationships being derived between flare plasma properties (temperature, emission measure etc.), which can in turn place constraints on the `allowed' values of properties for a flare of a given GOES magnitude.

TEBBS was successfully applied to 50,056 flares from B-class to X-class, making it the largest study of the thermal properties of solar flares to date.  It was found that peak temperature scales logarithmically with peak long channel flux as described by Equation~\ref{eqn:t(f)fit}.  Meanwhile, peak emission measure and total radiative losses scaled with peak long channel flux as power-laws given by Equations~\ref{eqn:em(f)fit} and \ref{eqn:lradfit}.  Uncertainties were calculated for these derived relations unlike previous studies.  The exception to this was \citet{batt05} who provided uncertainties for their slopes only.  The uncertainties derived using TEBBS were nonetheless smaller than those of \citet{batt05} and include uncertainties on the intercepts as well as slopes.  Furthermore, while these results are broadly in line with previous studies, it was found that flares of a given \goes class have lower temperatures and higher peak emission measures than previously reported.

Peak emission measure and total radiative losses were also examined as a function of peak temperature. It was found that flares with high peak temperatures also have high peak emission measures (in agreement with \citealt{garc88} and \citealt{garc92}). However, the derived correlation was relatively weak.  Similarly, it was also found that flares of a given peak temperature could exhibit a large range of radiative losses with no clearly defined trend.  This lack of a clearly defined relationship between two derived properties could be attributed to the assumptions that go in to calculating them.  Although both a constant density and a fixed coronal abundance were assumed in this study, both have been shown to vary during individual events \citep[e.g.][]{grah11}.  A followup analysis of how changes in these variables might affect the derived properties, particularly in conjunction with hydrodynamic simulations, may lead to more reasonable correlations.

This compilation of solar flare properties represents a valuable resource from which to conduct future large-scale statistical studies of flare plasma properties.  For example, \cite{stoi08} derived analytical predictions of temperature and emission measure in response to electron beam and conduction driven heating and compared the results to \rhessi observations of 18 microflares.  They found an order of magnitude discrepancy between conduction driven emission measures predicted by the Rosner-Tucker-Vaiana (RTV; \citealt{rtv78}) scaling laws and observation.  This seemed to suggest that electron beam processes dominated.  However, they noted that \rhessis's high temperature sensitivity ($\gtrsim$10~MK) mean that the observed temperatures may not have well represented the conduction value of the microflares, thus explaining the discrepancy.  The fact that the \goes XRS has a sensitivity to lower temperatures than \rhessi makes the TEBBS database ideal for exploring this possibility.  Since the RTV scaling laws and electron beam heating models are widely used to understand and model solar flares, it is important to examine disagreements between their predictions and observation.

Another example of the use of RTV scaling laws in understanding flares is \cite{asch08}.  They used these laws to derive theoretical ($EM \propto T^{4.3}$) and observed ($EM \propto T^{4.7}$) scaling laws between peak temperature and emission measure for solar and stellar flares.  However, as part of their study, results from previous studies such as \cite{feld96b} and \cite{feld95} were included which did not account for background issues.  TEBBS can therefore also be used to examine these scaling laws with greater statistical certainty and therefore provide more clarity on the discrepancies between theory and observation.  As the scaling laws derived by \cite{asch08} apply to solar and stellar flares, conclusions drawn from TEBBS can be extended to stellar flares as well.

TEBBS can be used to examine a wide range of flare characteristics, such as thermodynamic evolution and, in light of the work of \cite{stoi08}, even flare loop topologies.  As TEBBS is also the largest database of thermal flare plasma properties to date, it will provide a valuable resource for future solar flare research.

\acknowledgments
DFR would like to thank the Irish Research Council for Science, Engineering and Technology (IRCSET) for funding this research. ROM would like to thank Queen's University Belfast for the award of a Leverhulme Trust Research Fellowship. In addition, thanks are given to Drs. Stephen White, Jack Ireland, D. Shaun Bloomfield, and Claire L. Raftery for their insightful discussions which contributed to this work.

\appendix
\section{\goes saturation levels}
During the period of this study, there were 32 X-class flares in the \goes event list which saturated either the short channel or both channels. No events included in this study saturated the long channel without also saturating the short channel.  Saturation of the \goes channels has different and important effects when deriving each of the flare plasma properties. If the short channel saturates but the long channel does not, then the derived temperature during the period of saturation is a lower limit because $T \propto F_S / F_L$, to a first order approximation. However, derived emission measure during the same period is an upper limit because $EM \propto T^{-1}$, to a first order approximation. Likewise, since $dL_{rad}/dt \propto EM$, radiative loss rates (and total radiative losses) are also upper limits.  If both channels saturate however, then it cannot be determined (without extrapolation of the lightcurves) whether properties derived during the saturation period are upper or lower limits since they are all functions of the flux ratio.

Within the \goes event list used in this study, only the XRSs onboard \goess -6, \goess -10, and \goess -12 were seen to saturate.  Each channel in each XRS had different saturation levels which can be seen in Table~\ref{tab:satur}.  These values were taken from the \goes lightcurves of saturated events throughout the \goes event list.  \goess -6 had a very long lifetime and as a result, the saturation levels of each channel were seen to degrade over time, also shown in Table~\ref{tab:satur}.  (Dates are inclusive.)

\begin{table}
\begin{center}
\caption[\goes Saturation Levels]{\goes Saturation Levels}
\label{tab:satur}
\begin{tabular}{cccc}
\hline
\hline
\goes Satellite	&Time Period	&Long Channel Flux			&Short Channel Flux		\\
 			&			&(10$^{-4}$~W~m$^{-2}$) 	&(10$^{-4}$~W~m$^{-2}$)	\\
\hline
\goess -6		&06-Nov-1980~--~17-Dec-1982	&(...) 		&1.8 	\\
\goess -6		&24-Apr-1984                                	&13		&1.2	\\
\goess -6		&20-May-1984~--~24-Jun-1988	&(...) 		&1.2	\\
\goess -6		&06-Mar-1989~--~02-Nov-1992	&12		&1.2	\\
\goess -10	&02-Apr-2001~--~15-Apr-2001		&18		&4.7	\\
\goess -12	&28-Oct-2003~--~7-Sep-2005		&17		&4.9	\\
\hline
\end{tabular}
\end{center}
\end{table}

\bibliographystyle{apj}
\bibliography{ms}

\begin{thebibliography}{40}
\expandafter\ifx\csname natexlab\endcsname\relax\def\natexlab#1{#1}\fi

\bibitem[{{Aschwanden} {et~al.}(2008){Aschwanden}, {Stern}, \&
  {G{\"u}del}}]{asch08}
{Aschwanden}, M.~J., {Stern}, R.~A., \& {G{\"u}del}, M. 2008, \apj, 672, 659

\bibitem[{{Aschwanden} \& {Tsiklauri}(2009)}]{asch09}
{Aschwanden}, M.~J. \& {Tsiklauri}, D. 2009, \apjs, 185, 171

\bibitem[{{Battaglia} {et~al.}(2009){Battaglia}, {Fletcher}, \&
  {Benz}}]{batt09}
{Battaglia}, M., {Fletcher}, L., \& {Benz}, A.~O. 2009, \aap, 498, 891

\bibitem[{{Battaglia} {et~al.}(2005){Battaglia}, {Grigis}, \& {Benz}}]{batt05}
{Battaglia}, M., {Grigis}, P.~C., \& {Benz}, A.~O. 2005, \aap, 439, 737

\bibitem[{{Bhattacharya} {et~al.}(1983){Bhattacharya}, {Chernoff}, \&
  {Yang}}]{bhat83}
{Bhattacharya}, P.~K., {Chernoff}, H., \& {Yang}, S.~S. 1983, {Annals of
  Statistics}, 11, 505

\bibitem[{{Bornmann}(1990)}]{born90}
{Bornmann}, P.~L. 1990, \apj, 356, 733

\bibitem[{{Carmichael}(1964)}]{carm64}
{Carmichael}, H. 1964, NASA Special Publication, 50, 451

\bibitem[{{Caspi}(2010)}]{casp10}
{Caspi}, A. 2010, PhD thesis, Department of Physics, University of California,
  Berkeley, CA 94720-7450, USA

\bibitem[{{Christe} {et~al.}(2008){Christe}, {Hannah}, {Krucker}, {McTiernan},
  \& {Lin}}]{chri08}
{Christe}, S., {Hannah}, I.~G., {Krucker}, S., {McTiernan}, J., \& {Lin}, R.~P.
  2008, \apj, 677, 1385

\bibitem[{{Cox} \& {Tucker}(1969)}]{cox69}
{Cox}, D.~P. \& {Tucker}, W.~H. 1969, \apj, 157, 1157

\bibitem[{{Dere}(2007)}]{dere07}
{Dere}, K.~P. 2007, \aap, 466, 771

\bibitem[{{Dere} {et~al.}(2009){Dere}, {Landi}, {Young}, {Del Zanna},
  {Landini}, \& {Mason}}]{dere09}
{Dere}, K.~P., {Landi}, E., {Young}, P.~R., {Del Zanna}, G., {Landini}, M., \&
  {Mason}, H.~E. 2009, \aap, 498, 915

\bibitem[{{Feldman} {et~al.}(2004){Feldman}, {Dammasch}, {Landi}, \&
  {Doschek}}]{feld04}
{Feldman}, U., {Dammasch}, I., {Landi}, E., \& {Doschek}, G.~A. 2004, \apj,
  609, 439

\bibitem[{{Feldman} {et~al.}(1996{\natexlab{a}}){Feldman}, {Doschek}, \&
  {Behring}}]{feld96a}
{Feldman}, U., {Doschek}, G.~A., \& {Behring}, W.~E. 1996{\natexlab{a}}, \apj,
  461, 465

\bibitem[{{Feldman} {et~al.}(1996{\natexlab{b}}){Feldman}, {Doschek},
  {Behring}, \& {Phillips}}]{feld96b}
{Feldman}, U., {Doschek}, G.~A., {Behring}, W.~E., \& {Phillips}, K.~J.~H.
  1996{\natexlab{b}}, \apj, 460, 1034

\bibitem[{{Feldman} {et~al.}(1995){Feldman}, {Doschek}, {Mariska}, \&
  {Brown}}]{feld95}
{Feldman}, U., {Doschek}, G.~A., {Mariska}, J.~T., \& {Brown}, C.~M. 1995,
  \apj, 450, 441

\bibitem[{{Feldman} {et~al.}(1992){Feldman}, {Mandelbaum}, {Seely}, {Doschek},
  \& {Gursky}}]{feld92}
{Feldman}, U., {Mandelbaum}, P., {Seely}, J.~F., {Doschek}, G.~A., \& {Gursky},
  H. 1992, \apjs, 81, 387

\bibitem[{{Fisher} {et~al.}(1985){Fisher}, {Canfield}, \& {McClymont}}]{fisc85}
{Fisher}, G.~H., {Canfield}, R.~C., \& {McClymont}, A.~N. 1985, \apj, 289, 414

\bibitem[{{Fletcher} {et~al.}(2011){Fletcher}, {Dennis}, {Hudson}, {Krucker},
  {Phillips}, {Veronig}, {Battaglia}, {Bone}, {Caspi}, {Chen}, {Gallagher},
  {Grigis}, {Ji}, {Liu}, {Milligan}, \& {Temmer}}]{flet11}
{Fletcher}, L., {Dennis}, B.~R., {Hudson}, H.~S., {Krucker}, S., {Phillips},
  K., {Veronig}, A., {Battaglia}, M., {Bone}, L., {Caspi}, A., {Chen}, Q.,
  {Gallagher}, P., {Grigis}, P.~T., {Ji}, H., {Liu}, W., {Milligan}, R.~O., \&
  {Temmer}, M. 2011, \ssr, 261

\bibitem[{{Fludra} {et~al.}(1995){Fludra}, {Doyle}, {Metcalf}, {Lemen},
  {Phillips}, {Culhane}, \& {Kosugi}}]{flud95}
{Fludra}, A., {Doyle}, J.~G., {Metcalf}, T., {Lemen}, J.~R., {Phillips},
  K.~J.~H., {Culhane}, J.~L., \& {Kosugi}, T. 1995, \aap, 303, 914

\bibitem[{{Garcia}(1988)}]{garc88}
{Garcia}, H.~A. 1988, Advances in Space Research, 8, 157

\bibitem[{{Garcia}(1994)}]{garc94}
---. 1994, \solphys, 154, 275

\bibitem[{{Garcia} \& {McIntosh}(1992)}]{garc92}
{Garcia}, H.~A. \& {McIntosh}, P.~S. 1992, \solphys, 141, 109

\bibitem[{{Graham} {et~al.}(2011){Graham}, {Fletcher}, \& {Hannah}}]{grah11}
{Graham}, D.~R., {Fletcher}, L., \& {Hannah}, I.~G. 2011, \aap, 532, A27

\bibitem[{{Hannah} {et~al.}(2008){Hannah}, {Christe}, {Krucker}, {Hurford},
  {Hudson}, \& {Lin}}]{hann08}
{Hannah}, I.~G., {Christe}, S., {Krucker}, S., {Hurford}, G.~J., {Hudson},
  H.~S., \& {Lin}, R.~P. 2008, \apj, 677, 704

\bibitem[{{Hanser} \& {Sellers}(1996)}]{hans96}
{Hanser}, F.~A. \& {Sellers}, F.~B. 1996, in Society of Photo-Optical
  Instrumentation Engineers (SPIE) Conference, Vol. 2812, Society of
  Photo-Optical Instrumentation Engineers (SPIE) Conference Series, ed.
  {E.~R.~Washwell}, 344--352

\bibitem[{{Hirayama}(1974)}]{hira74}
{Hirayama}, T. 1974, \solphys, 34, 323

\bibitem[{{Kopp} \& {Pneuman}(1976)}]{kopp76}
{Kopp}, R.~A. \& {Pneuman}, G.~W. 1976, \solphys, 50, 85

\bibitem[{{Landi} {et~al.}(2002){Landi}, {Feldman}, \& {Dere}}]{land02}
{Landi}, E., {Feldman}, U., \& {Dere}, K.~P. 2002, \apjs, 139, 281

\bibitem[{{Landi} {et~al.}(1999){Landi}, {Landini}, {Dere}, {Young}, \&
  {Mason}}]{land99}
{Landi}, E., {Landini}, M., {Dere}, K.~P., {Young}, P.~R., \& {Mason}, H.~E.
  1999, \aaps, 135, 339

\bibitem[{{Lin} {et~al.}(2002){Lin}, {Dennis}, {Hurford}, {Smith}, {Zehnder},
  {Harvey}, {Curtis}, {Pankow}, {Turin}, {Bester}, {Csillaghy}, {Lewis},
  {Madden}, {van Beek}, {Appleby}, {Raudorf}, {McTiernan}, {Ramaty}, {Schmahl},
  {Schwartz}, {Krucker}, {Abiad}, {Quinn}, {Berg}, {Hashii}, {Sterling},
  {Jackson}, {Pratt}, {Campbell}, {Malone}, {Landis}, {Barrington-Leigh},
  {Slassi-Sennou}, {Cork}, {Clark}, {Amato}, {Orwig}, {Boyle}, {Banks},
  {Shirey}, {Tolbert}, {Zarro}, {Snow}, {Thomsen}, {Henneck}, {McHedlishvili},
  {Ming}, {Fivian}, {Jordan}, {Wanner}, {Crubb}, {Preble}, {Matranga}, {Benz},
  {Hudson}, {Canfield}, {Holman}, {Crannell}, {Kosugi}, {Emslie}, {Vilmer},
  {Brown}, {Johns-Krull}, {Aschwanden}, {Metcalf}, \& {Conway}}]{lin02}
{Lin}, R.~P., {Dennis}, B.~R., {Hurford}, G.~J., {Smith}, D.~M., {Zehnder}, A.,
  {Harvey}, P.~R., {Curtis}, D.~W., {Pankow}, D., {Turin}, P., {Bester}, M.,
  {Csillaghy}, A., {Lewis}, M., {Madden}, N., {van Beek}, H.~F., {Appleby}, M.,
  {Raudorf}, T., {McTiernan}, J., {Ramaty}, R., {Schmahl}, E., {Schwartz}, R.,
  {Krucker}, S., {Abiad}, R., {Quinn}, T., {Berg}, P., {Hashii}, M.,
  {Sterling}, R., {Jackson}, R., {Pratt}, R., {Campbell}, R.~D., {Malone}, D.,
  {Landis}, D., {Barrington-Leigh}, C.~P., {Slassi-Sennou}, S., {Cork}, C.,
  {Clark}, D., {Amato}, D., {Orwig}, L., {Boyle}, R., {Banks}, I.~S., {Shirey},
  K., {Tolbert}, A.~K., {Zarro}, D., {Snow}, F., {Thomsen}, K., {Henneck}, R.,
  {McHedlishvili}, A., {Ming}, P., {Fivian}, M., {Jordan}, J., {Wanner}, R.,
  {Crubb}, J., {Preble}, J., {Matranga}, M., {Benz}, A., {Hudson}, H.,
  {Canfield}, R.~C., {Holman}, G.~D., {Crannell}, C., {Kosugi}, T., {Emslie},
  A.~G., {Vilmer}, N., {Brown}, J.~C., {Johns-Krull}, C., {Aschwanden}, M.,
  {Metcalf}, T., \& {Conway}, A. 2002, \solphys, 210, 3

\bibitem[{{Mazzotta} {et~al.}(1998){Mazzotta}, {Mazzitelli}, {Colafrancesco},
  \& {Vittorio}}]{mazz98}
{Mazzotta}, P., {Mazzitelli}, G., {Colafrancesco}, S., \& {Vittorio}, N. 1998,
  \aaps, 133, 403

\bibitem[{{Phillips} \& {Feldman}(1995)}]{phil95}
{Phillips}, K.~J.~H. \& {Feldman}, U. 1995, \aap, 304, 563

\bibitem[{{Phillips} {et~al.}(2010){Phillips}, {Sylwester}, {Sylwester}, \&
  {Kuznetsov}}]{phil10}
{Phillips}, K.~J.~H., {Sylwester}, J., {Sylwester}, B., \& {Kuznetsov}, V.~D.
  2010, \apj, 711, 179

\bibitem[{{Rosner} {et~al.}(1978){Rosner}, {Tucker}, \& {Vaiana}}]{rtv78}
{Rosner}, R., {Tucker}, W.~H., \& {Vaiana}, G.~S. 1978, \apj, 220, 643

\bibitem[{{Rousseeuw}(1984)}]{rous84}
{Rousseeuw}, P.~J. 1984, {Journal of the American Statistical Association}, 79,
  871

\bibitem[{{Stoiser} {et~al.}(2008){Stoiser}, {Brown}, \& {Veronig}}]{stoi08}
{Stoiser}, S., {Brown}, J.~C., \& {Veronig}, A.~M. 2008, \solphys, 250, 315

\bibitem[{{Sturrock}(1966)}]{stur66}
{Sturrock}, P.~A. 1966, \nat, 211, 695

\bibitem[{{Thomas} {et~al.}(1985){Thomas}, {Crannell}, \& {Starr}}]{thom85}
{Thomas}, R.~J., {Crannell}, C.~J., \& {Starr}, R. 1985, \solphys, 95, 323

\bibitem[{{White} {et~al.}(2005){White}, {Thomas}, \& {Schwartz}}]{whit05}
{White}, S.~M., {Thomas}, R.~J., \& {Schwartz}, R.~A. 2005, \solphys, 227, 231

\end{thebibliography}

\end{document}